\documentclass[twocolumn,showpacs,preprintnumbers,amsmath,amssymb]{revtex4}

\newcommand{\bib}{\bibitem}
\newcommand{\bea}{\begin{eqnarray}}
\newcommand{\eea}{\end{eqnarray}}
\newcommand{\beq}{\begin{equation}}
\newcommand{\eeq}{\end{equation}}
\newcommand{\non}{\nonumber}
\newcommand{\da}{\dagger}
\newcommand{\al}{\alpha}

\newcommand{\ga}{\gamma}
\newcommand{\de}{\delta}
\newcommand{\ep}{\epsilon}
\newcommand{\om}{\omega}
\newcommand{\si}{\sigma}

\usepackage{graphicx,epsfig} % Include figure files
\usepackage{dcolumn} % Align table columns on decimal point
\usepackage{bm} % bold math

\begin{document}

\title{Charge transport in a Tomonaga-Luttinger liquid: effects of 
pumping and bias}
\author{Amit Agarwal and Diptiman Sen}
\affiliation{Center for High Energy Physics, Indian Institute of Science,
Bangalore 560012, India}
\date{\today}

\begin{abstract}
We study the current produced in a Tomonaga-Luttinger liquid by an applied 
bias and by weak, point-like impurity potentials which are oscillating in 
time. We use bosonization to perturbatively calculate the current up to 
second order in the impurity potentials. In the regime of small bias and 
low pumping frequency, both the DC and AC components of the current have 
power law dependences on the bias and pumping frequencies with an exponent 
$2K-1$ for spinless electrons, where $K$ is the interaction parameter. For 
$K < 1/2$, the current grows large for special values of the bias. For 
non-interacting electrons with $K=1$, our results agree with those obtained 
using Floquet scattering theory for Dirac fermions. We also discuss the cases 
of extended impurities and of spin-1/2 electrons.
\end{abstract}

\pacs{73.23.-b, 73.63.Nm, 71.10.Pm}
\maketitle

\section{Introduction}

The conductance of electrons in a quantum wire has been studied extensively
in recent years both theoretically \cite{datta,imry} and experimentally 
\cite{tarucha,liang,bkane,yacoby,auslaender,reilly}. For a wire in which only
one channel is available to the electrons and the transport is ballistic 
(i.e., there are no impurities inside the wire, and there is no scattering 
from phonons or from the contacts between the wire and its leads), the 
conductance is given by $G = 2e^2 /h$ for infinitesimal bias. However, if 
there is an impurity inside the wire which scatters the electrons, then the 
conductance is reduced. For a $\delta$-function impurity with strength $U$, we
obtain $G = (2e^2 /h)~ (1 - U^2 /v_F^2),$ to lowest order in $U$, where $v_F$
is the Fermi velocity of the electrons. In the presence of 
interactions between the electrons, the impurity strength $U$ effectively 
becomes a function of the length scale through a renormalization group (RG) 
equation \cite{kane,furusaki}. The RG flow has to be cut-off at the smallest 
of the three length scales of the system, namely, the wire length, the thermal
length which is inversely proportional to the temperature, and a length which 
is inversely proportional to the bias voltage $V_{bias}$. If the latter length
scale is the smallest of the three, then the effective value of the impurity 
strength is given by $U V_{bias}^{K-1}$, where $K$ is a parameter related to 
the strength of the interactions between the electrons as we will see later. 
Hence the correction to the conductance due to the combined effect of the 
impurity and the interactions is given by $\Delta G \sim U^2 V_{bias}^{2K-2}$.

The phenomenon of charge pumping and rectification by oscillating potentials
applied to certain points in a system has also been studied theoretically 
[11-38] and experimentally \cite{switkes,taly1,cunningham,taly2,leek}. 
For the case of non-interacting electrons, theoretical studies have used 
adiabatic scattering theory \cite{avron,entin1,entin2}, Floquet scattering 
theory \cite{moskalets,kim}, variations of the non-equilibrium Green function 
formalism \cite{wang,arrachea,torres}, and the equation of motion approach 
\cite{agarwal}. The case of interacting electrons has also been studied, using
a RG method for weak interactions \cite{das}, and the method of bosonization 
for arbitrary interactions [45-55]. The analytical methods used in the two 
cases typically treat the electrons in quite different ways, with a 
non-relativistic Schr\"odinger equation or a tight-binding
model on a lattice being used in the non-interacting case, and a massless 
Dirac model followed by bosonization being used in the interacting case.

A clear comparison between the non-interacting and interacting cases does not 
seem to have been made before. We plan to fill this gap in this paper and 
will study the effects of electron-electron interactions (of arbitrary 
repulsive strength) on the DC and AC components of the current in a system 
with a bias and time-dependent impurities. 
 
In Sec. II, we discuss a massless Dirac model for non-interacting 
electrons in the presence of several point-like impurities. We use 
Floquet scattering theory \cite{moskalets,kim} to study the pumped and bias
current in this model. In Sec. III, we first review the calculation of the
backscattered current. Using bosonization \cite{gogolin,rao,giamarchi}, we 
then compute the DC and AC components of the current up to second order in 
the impurity potentials \cite{feldman1,makogon1}. We show that these reduce to
the results obtained using Floquet scattering theory for the non-interacting 
case. The cases of extended impurities \cite{makogon2} and of spin-1/2 
electrons are discussed briefly. We summarize our results in Sec. IV.
We will consider infinitely long wires and zero temperature throughout the 
paper; hence the relevant energy scales in the problem are set only by the 
bias and the pumping frequency.

\section{Massless Dirac model with impurities}

In this section we consider a spinless and massless Dirac fermion with no 
interactions between the fermions, and use the Floquet scattering theory
to compute the total current. The Hamiltonian in the presence of several 
point-like impurities is given by $\hat H = \hat H_0 + \hat H_{imp}$, where
\bea 
\hat H_0 &=& \int dx ~i v_F ~(- ~\psi_R^\da \frac{\partial \psi_R}{\partial 
x} ~ +~ \psi^\da_L \frac{\partial \psi_L}{\partial x}) , \non \\
\hat H_{imp} &=& \int dx ~\sum_p ~\de (x-x_p) ~U_p(t) ~\psi^\dag (x) \psi 
(x), \non \\
\psi (x) &=& \psi_R (x) ~e^{ik_F x} ~+~ \psi_L (x) ~
e^{-ik_F x} , \label{ham} \eea
where $\psi_L$ and $\psi_R$ are the fermionic field operators of the left and 
right moving electrons, $v_F$ is the Fermi velocity, and $k_F$ is the Fermi
wavenumber which originates from some underlying microscopic model. For 
instance, one may have a system of non-relativistic electrons with a Fermi
energy $E_F = k_F^2 /(2m)$ and $v_F = k_F /m$. (We are setting Planck's 
constant $\hbar$ equal to unity). The Hamiltonian $\hat H_0$ is obtained by 
linearizing the dispersion around the two Fermi points given by $k=\pm k_F$. 
$\hat H_{imp}$ arises from the time-dependent impurities which have strengths 
$U_p(t)$; this Hamiltonian couples left and right moving fields since
\beq \psi^\dag \psi ~=~ \psi_R^\dag \psi_R ~+~ \psi_L^\dag \psi_L ~+~ 
\psi_R^\dag \psi_L e^{-i2k_F x} ~+~ \psi_L^\dag \psi_R e^{i2k_F x}. 
\label{density} \eeq
We will assume that 
\beq U_p (t) ~=~ U_p ~\cos (\om t + \phi_p), \label{upt} \eeq
i.e., all impurities vary harmonically in time with the same frequency $\om$.
We will now use Floquet scattering theory and carry out a perturbative 
expansion in the dimensionless quantities $U_p/ v_F$.

The equations of motion in the presence of a single $\de$-function 
impurity $\de (x-x_p) U_p \cos (\om t + \phi_p)$ is as follows:
\bea & & i~ \frac{\partial \psi_R}{\partial t} ~+~ i v_F ~ \frac{\partial 
\psi_R}{\partial x} \non \\
& & =~ \de (x-x_p) ~U_p \cos (\om t + \phi_p) ~(\psi_R ~+~ \psi_L 
e^{-i2k_F x_p}), \non \\
& & i~ \frac{\partial \psi_L}{\partial t} ~-~ i v_F ~\frac{\partial 
\psi_L}{\partial x} \non \\
& & =~ \de (x-x_p) ~U_p \cos (\om t + \phi_p) ~(\psi_L ~+~ \psi_R 
e^{i2k_F x_p}). \label{eom} \non \\
& & \eea
If we define the linear combinations $\psi_+ = \psi_R e^{ik_F x_p} + \psi_L
e^{-ik_F x_p}$ and $\psi_- = \psi_R e^{ik_F x_p} - \psi_L e^{-ik_F x_p}$, we 
find that
\bea & & i~ \frac{\partial \psi_-}{\partial t} ~+~ i v_F ~\frac{\partial
\psi_+}{\partial x} = 0 , \non \\
& & i~ \frac{\partial \psi_+}{\partial t} ~+~ i v_F ~\frac{\partial \psi_-}{
\partial x} = 2 \de (x-x_p) U_p \cos (\om t + \phi_p) ~\psi_+ . \non \\
& & \eea
By integrating over a little region from $x_p - \ep$ to $x_p + \ep$, we
find that $\psi_+$ is continuous at the point $x=x_p$, while $\psi_-$ has
a discontinuity given by
\beq iv_F ~[\psi_- (x_p + \ep) ~-~ \psi_- (x_p - \ep)] ~=~ 2 U_p \cos 
(\om t + \phi_p) ~\psi_+ (x_p). \label{bc} \eeq
We would like to note here that it is necessary to retain the terms
$\psi_R^\dag \psi_R + \psi_L^\dag \psi_L$ in Eq. (\ref{density}) in order
to have continuity of $\psi_+$. In some papers, these terms are not taken
into consideration. One then runs into the mathematical peculiarity
that $\psi_+$ and $\psi_-$ are both discontinuous at $x=x_p$, and the
discontinuity is taken to be proportional to their values at that point;
but those values are actually ill-defined due to the discontinuity.

We can now solve Eqs. (\ref{eom}) along with the boundary conditions in Eq. 
(\ref{bc}). For a single $\de$-function impurity oscillating with frequency
$\om$ at $x=x_p$, let us consider a wave coming from the left ($x < x_p$) 
with energy $E_0$ and unit amplitude. Note that we are measuring energies 
with respect to a Fermi energy, so that $E_0 =0$ corresponds to a fermion at 
the Fermi energy. Due to the oscillating impurity potential, the wave will be 
reflected back to the left with energy $E_n \equiv E_0 +n\om$ and 
amplitude $S_{LL} (E_n , E_0)$, or transmitted to the right ($x > x_p$) with
energy $E_n$ and amplitude $S_{RL} (E_n , E_0)$, where $n = 0, \pm 1, \pm 2, 
\cdots$ defines the Floquet side bands \cite{moskalets}. Note that since we 
are considering a Dirac fermion, there is no upper or lower bound to the 
energy $E_n$, and the velocity $v_F$ is independent of the energy. (This is 
unlike the case of a non-relativistic fermion or a fermion on a lattice 
where there is a lower or upper bound to the energy, and the velocity is 
a function of the energy). To be explicit, the wave function is given by
\bea \psi_R &=& e^{i(k_0x-E_0t)} \quad {\rm for} \quad x < x_p ~, \non \\
&=& \sum_n ~S_{RL} (E_n , E_0) ~e^{i(k_nx-E_nt)} \quad {\rm for} \quad
x > x_p ~, \non \\
\psi_L &=& \sum_n ~S_{LL} (E_n , E_0) ~e^{i(-k_nx-E_nt)} \quad {\rm for} \quad
x < x_p ~, \non \\
&=& 0 \quad {\rm for} \quad x > x_p ~, \eea
where $k_n = E_n /v_F$. Similarly, we can consider a wave coming from the 
right with energy $E_0$ and unit amplitude; it will be reflected back to the 
right with amplitude $S_{RR} (E_n , E_0)$ or transmitted to the left with 
amplitude $S_{LR} (E_n , E_0)$. Let us simplify the notation by defining
\bea r_{L,n} &=& S_{LL} (E_n , E_0), \quad t_{L,n} ~=~ S_{LR} (E_n , E_0), 
\non \\
t_{R,n} &=& S_{RL} (E_n , E_0), \quad r_{R,n} ~=~ S_{RR} (E_n , E_0). \eea
Due to unitarity, we have the relations
\bea \sum_n ~[~ |r_{L,n}|^2 ~+~ |t_{R,n}|^2 ~] &=& 1, \non \\
\sum_n ~[~ |r_{R,n}|^2 ~+~ |t_{L,n}|^2 ~] &=& 1. \label{unit} \eea

The different Floquet scattering amplitudes $r_{\al,n}$ and $t_{\al,n}$ can 
be found by using the boundary conditions in Eq. (\ref{bc}). We will consider 
the case of several impurities labeled by the index $p$ as in Eq. (\ref{ham}).
To simplify our calculations, we will assume that $\om (x_p -x_r) /v_F$ and 
$E_0 (x_p - x_r)/v_F$ are small for all pairs of impurities $p$ and $r$; the 
first condition corresponds to the adiabatic limit, while the second condition
implies that we are only considering states close to the Fermi energy. Keeping
terms only up to first order in $U_p /v_F$, we find that only the first
Floquet side bands are excited, and
\bea 
t_{L,1} &=& t_{R,1} ~=~ - ~\frac{i}{2v_F} ~\sum_p ~U_p ~e^{-i\phi_p}, \non \\
t_{L,-1} &=& t_{R,-1} ~=~ -~\frac{i}{2v_F} ~\sum_p ~U_p ~e^{i\phi_p}. \non \\
r_{L,1} &=& -~ \frac{i}{2v_F} ~\sum_p ~U_p ~e^{i(2k_F x_p -\phi_p)}, \non \\
r_{L,-1} &=& -~ \frac{i}{2v_F} ~\sum_p ~U_p ~e^{i(2k_F x_p +\phi_p)}. \non \\
r_{R,1} &=& -~ \frac{i}{2v_F} ~\sum_p ~U_p ~e^{i(-2k_F x_p -\phi_p)}, \non \\
r_{R,-1} &=& -~ \frac{i}{2v_F} ~\sum_p ~U_p ~e^{i(-2k_F x_p +\phi_p)}, 
\label{tr} \eea
We also find that the unitarity relations in Eq. (\ref{unit}) are satisfied 
up to second order in $U_p /v_F$, and therefore $t_{L,0}$ and $t_{R,0}$ are 
given by 
\bea |t_{L,0}|^2 &=& 1 - |r_{R,1}|^2 - |r_{R,-1}|^2 - |t_{L,1}|^2 -
|t_{L,-1}|^2, \non \\
|t_{R,0}|^2 &=& 1 - |r_{L,1}|^2 - |r_{L,-1}|^2 - |t_{R,1}|^2 - 
|t_{R,-1}|^2, \non \\
& & \label{t0} \eea
to that order in $U_p /v_F$. Note that the amplitudes given in Eqs. 
(\ref{tr}-\ref{t0}) are all independent of $E_0$ under the approximations 
that we have made.

The dc part of the current in, say, the right lead is given by \cite{moskalets}
\bea I_{R,dc} &=& q ~\int_{-\infty}^\infty \frac{dE_0}{2\pi} ~\sum_n ~[~
|r_{R,n}|^2 ~\{f_R (E_0) - f_R (E_n)\} \non \\
& & \quad \quad \quad \quad \quad \quad ~+~ |t_{R,n}|^2 ~\{f_L (E_0) - f_R 
(E_n)\} ~], \non \\
& & \label{irdc} \eea
where $q$ is the charge of the electron, and $f_\al (E)=1/[e^{(E- 
\mu_\al)/k_BT} + 1]$ is the Fermi function in the lead $\al$. At zero 
temperature, $f_\al (E) = \theta (\mu_\al - E)$, where $\theta (E) = 1$ for 
$E>0$ and 0 for $E<0$. Note that in a non-relativistic system, an expression 
like (\ref{irdc}) would contain ratios of velocities $v_n/v_0$. Since we are 
considering a massless Dirac fermion here, the velocity is independent of the 
energy, and $v_n/v_0 =1$ for all $n$.

Let us define a frequency in terms of the bias voltage,
$\om_0 = q V_{bias} = \mu_R -\mu_L$. We will assume the bias to be small so 
that $\om_0 (x_r - x_p)/v_F$ is small for all values of $p$ and $r$. Using
Eqs. (\ref{tr}-\ref{t0}), we find, up to second order in $U_p /v_F$, that
\bea I_{R,dc} &=& - ~\frac{q \om_0}{2\pi} ~[~ |t_{R,0}|^2 ~+~ |t_{R,1}|^2 ~
+~ |t_{R,-1}|^2 ~] \non \\
& & + ~\frac{q\om}{2\pi} ~[~ |t_{R,1}|^2 - |t_{R,-1}|^2 + |r_{R,1}|^2 - 
|r_{R,-1}|^2 ~] \non \\
&=& -~ \frac{q \om_0}{2\pi} ~+~ \frac{q \om_0}{4\pi v_F^2} ~[\sum_p ~U_p^2 
\non \\
& & \quad \quad \quad \quad ~+~ 2~ \sum_{p<r} ~U_p U_r \cos(2k_F x_{rp}) 
\cos (\phi_{rp})] \non \\
& & - ~\frac{q\om}{2\pi v_F^2} ~ \sum_{p<r} ~U_p U_r ~ \sin(2k_F x_{rp}) 
\sin (\phi_{rp}), \label{ir} \eea
where $x_{rp} = x_r - x_p$ and $\phi_{rp} = \phi_r - \phi_p$. Eq. (\ref{ir})
shows the effects of a bias ($\om_0$) and harmonically oscillating 
potentials ($\om$). For the pure pumping case with $\om_0=0$, Eq. (\ref{ir})
agrees with the results presented in Ref. \cite{agarwal}; note that the
pumped current depends on $\sin (\phi_{rp})$.

It is interesting to note that the first term is just the ballistic 
conductance of a clean wire multiplied by the bias, the second term is a
correction to the clean case because of the presence of impurities, and the 
third term is the pumped current. In the non-interacting case, the bias 
component and the pumped component separate out, but for the interacting 
case, the current involves powers of $\om_0 \pm \om $. 

\section{Bosonization calculation of backscattered current}

\subsection{Backscattering current operator}

We now compute the current in a system of interacting 
electrons using the backscattering current operator introduced in Refs.
\cite{chamon1,chamon2,sharma,feldman1}.

Let us take the impurity potentials to be absent at time $t=- \infty$; then 
they are gradually switched on. At the initial time, $H_0$ commutes with the 
number operators of the left moving and right moving
fermions, $\hat N_L$ and $\hat N_R$ respectively. In the absence of any
impurity potentials, all the right movers originate in the left reservoir 
which is maintained at the chemical potential $\mu_L$, and all the left movers 
originate in the right reservoir maintained at the chemical potential $\mu_R$.
Hence, the system is initially described in the grand canonical ensemble by 
the chemical potentials $\mu_L$ and $\mu_R$ which are the coefficients of 
the number operators $\hat N_L$ and $\hat N_R$ respectively. 
We will work in the interaction representation, taking the chemical
potentials to be part of the interaction. This introduces time 
dependences into the fermionic operators $\psi_L \to \psi_L e^{i\mu_Lt}$ and
and $\psi_R \to \psi_R e^{i\mu_Rt}$. The operators $\psi_L^\dag \psi_R$ and
$\psi_R^\dag \psi_L$ appearing in in $H_{imp}$ (see Eqs. (\ref{ham}) and
(\ref{density})) therefore pick up factors of $e^{\pm i \om_0 t}$.

If there were no impurities, there would be a current flowing to the left
given by $I_0 = q^2 V_{bias} /(2\pi) = q \om_0 /(2\pi)$. In the presence of 
impurities, some of this current is backscattered to the right. The total 
current flowing to the right is given by 
$I = - I_0 + I_{bs}$, where $I_{bs}$ is the correction to the current due to 
backscattering by the impurities. The backscattered current is defined as
\bea \hat I_{bs} (t) &=& q ~\frac{d\hat N_R}{dt} ~=~ -iq ~[\hat N_R ~,~ \hat
H_{imp}~] \non \\
&=& iq ~\sum_p ~U_p (t) ~[~\psi^\da_L \psi_R~ e^{i(\om_0 t + 2 k_F x_p)} 
- H.c.~]. \non \\
& & \eea

The backscattered current at any time $t$ is given by
\beq \langle \hat I_{bs} \rangle = \langle 0| ~S(-\infty;t) ~\hat I_{bs} (t)~
S(t;-\infty) ~|0 \rangle , \eeq
where $|0 \rangle$ denotes the initial state at $t \to -\infty$, and 
$S$ is the scattering matrix arising due to the impurities,
\beq S(t;-\infty) ~=~ S^\dag (-\infty;t) ~=~ T ~\exp [-i ~\int_{-\infty}^{t}
dt' ~\hat H_{imp}(t') ]. \eeq 
We define a backscattering operator
\beq \hat B_p(t) ~=~ U_p (t) ~\psi^\da_L \psi_R ~e^{i(\om_0 t + 2 k_F x_p)}. 
\eeq 
In terms of this, $\hat I_{bs} (t) = i q \sum_p [\hat B_p (t) - H.c.]$, while
$S (t,-\infty) = 1 - i \sum_p \int_{-\infty}^t dt' [\hat B_p (t') + H.c.]$ to
first order in $U_p$. Thus 
\bea I_{bs} &=& q ~\sum_{p,r} ~\int_{-\infty}^t dt' ~\langle 0|~ [~ \hat 
B_p(t') \hat B_r^\da (t) - \hat B_p^\da (t') \hat B_r(t) ~] \non \\
& & \quad \quad \quad \quad \quad \quad \quad \quad \quad +~ H.c. ~|0 \rangle
\label{ibs} \eea
to second order in $U_p$.

\subsection{Bosonization}

In one dimension, it is known that a large class of fermion systems which are 
gapless and have a low-energy dispersion which is linear can be written in 
terms of gapless bosonic systems \cite{gogolin,rao,giamarchi}. These systems 
are called Tomonaga-Luttinger liquids; for spinless fermions, they are 
characterized by an interaction parameter $K$ and a velocity $v$. 
Non-interacting fermions have $K=1$ and $v=v_F$, while $K<1$ corresponds to 
repulsive interactions between the fermions. To be specific, consider a 
system with short-range density-density interactions of the form 
\beq H_{int} ~=~ \frac{1}{2} ~\int \int ~dx dy ~\rho (x) V(x-y) \rho (y), 
\label{int} \eeq
where $V(x)$ is a real and even function of $x$, and the density $\rho =
\psi^\dag \psi$ is given in Eq. (\ref{density}). We can write Eq. (\ref{int}) 
in a simple way if $V(x)$ is so short ranged that the arguments $x$ and $y$ of
the two density fields can be set equal to each other wherever possible. Using
the anticommutation relations between the fermion fields, we obtain
\beq H_{int} ~=~ g_2 ~\int ~dx ~\psi_R^\dag \psi_R \psi_L^\dag \psi_L , \eeq
where $g_2$ is related to the Fourier transform of $V(x)$ as $g_2 = {\tilde V}
(0) - {\tilde V} (2k_F)$. Defining a parameter $\ga = g_2 /(2 \pi v_F)$, we 
have the relations
\bea K &=& \left( \frac{1 - \ga}{1 + \ga} \right)^{1/2} , \non \\
v &=& v_F ~(1 - \ga^2)^{1/2} . \eea

In the absence of impurities, the bosonic action is given by
\beq S ~=~ \int dt dx \left[ ~\frac{1}{2v} ~\left( \frac{\partial 
\phi}{\partial t} \right)^2 ~-~ \frac{v}{2} ~\left( \frac{\partial 
\phi}{\partial x} \right)^2 ~\right]. \eeq
Bilinears in fermion operators can be written in terms of bosons and Klein 
operators \cite{gogolin,rao,giamarchi}, such as
\bea \hat{\psi}_R^\dag \hat{\psi}_L &=& \frac{1}{2\pi \al} ~\hat{\eta}_R^\dag
\hat{\eta}_L ~e^{i2\sqrt{\pi K} \hat{\phi}}, \non \\
\hat{\psi}_L^\dag \hat{\psi}_R &=& \frac{1}{2\pi \al} ~\hat{\eta}_L^\dag
\hat{\eta}_R ~e^{-i2\sqrt{\pi K} \hat{\phi}}, \eea
where $\eta_R$ and $\eta_L$ are the Klein operators, and $\al$ is a short
distance cut-off. We then obtain the ground state expectation value
of products of four fermion operators as in Eq. (\ref{ibs}), namely,
\bea & & \langle 0| ~\psi_R^\da (x_p,t') \psi_L (x_p,t') \psi_L^\da (x_r,t) 
\psi_R (x_r,t) ~|0 \rangle \non \\
&=& \frac{\al^{2K-2}}{(2 \pi)^2 ~[(x_p-x_r)^2 ~-~ (v(t'-t) - i\al)^2]^K}
\label{expval1} \eea
for all values of $K$. 

For the non-interacting case with $K =1$, we can evaluate the above ground 
state expectation value directly without using bosonization. We use the
second quantized expressions for the fermion fields, 
\bea \psi_R &=& \int_{-\infty}^\infty ~\frac{dk}{2\pi} ~a_{Rk} ~e^{ik(x-v_Ft)},
\non \\
\psi_L &=& \int_{-\infty}^\infty ~\frac{dk}{2\pi} ~a_{Lk} ~e^{ik(-x-v_Ft)}, 
\eea
where the creation and annihilation operators satisfy the anticommutation
relations $\{ a_{Rk}, a_{Rk'}^\dag \} = \{ a_{Lk}, a_{Lk'}^\dag \} = 2\pi
\de (k-k')$. The ground state $|0\rangle$ is annihilated by $a_{Rk}, a_{Lk}$
for $k>0$ and by $a_{Rk}^\dag, a_{Lk}^\dag$ for $k<0$. We then find that the 
ground state expectation value agrees with the result given in Eq. 
(\ref{expval1}) for $K=1$ and $v=v_F$.

In general, the backscattered current has two parts: one independent of time 
which we call $I_{dc}$, and the other varying with time, with frequency 
$2 \om$ to second order in $U_p$, which we call $I_{ac}$. $I_{ac}$ does not 
contribute to any charge transfer as its average over a cycle is zero. In the 
next few subsections, we calculate the expectation value of the backscattered 
current for various cases and study them in different limits. To simplify our 
calculations, we again assume that $\om x_{rp}/v$ and $\om_0 x_{rp}/v$ are 
small and that $\om \ge 0$. It will be convenient to define 
the combinations
\beq \om_+ ~=~ \om_0 ~+~ \om , \quad {\rm and} \quad \om_- ~=~ \om_0 ~-~ \om.
\eeq

\subsection{Single impurity}

This case has been discussed in Ref. \cite{feldman1}; we reproduce the results
here for the sake of completeness. Some details of the calculations are 
provided in the Appendix.

\bea I_{bs,dc}^{pp} &=& \frac{qU_p^2}{8 \pi v^2 \Gamma (2K)} ~\left( 
\frac{\al}{v} \right)^{2K-2} \non \\ 
& \times & [sgn (\om_+) ~|\om_+|^{2K-1} ~+~ sgn (\om_-) ~|\om_-|^{2K-1}],
\non \\
& & \label{idc1} \eea
\bea I_{bs,ac}^{pp} &=& \frac{qU_p^2}{8 \pi v^2 \Gamma (2K) \cos (\pi K)} ~
\left(\frac{\al}{v} \right)^{2K-2} \non \\
&\times& [sgn (\om_+) ~|\om_+|^{2K-1} \non \\
& & ~\times ~\cos(2 \om t + 2 \phi_p + sgn (\om_+) \pi K) \non \\
& & ~+~ sgn (\om_-) ~|\om_-|^{2K-1} \non \\
& & ~\times ~\cos(2 \om t + 2\phi_p - sgn (\om_-) \pi K)], \non \\
& & \label{iac1} \eea
where $sgn(\Omega) \equiv 1$ if $\Omega > 0$, $0$ if $\Omega = 0$ and $-1$ if 
$\Omega < 0$. In Eqs. (\ref{idc1}-\ref{iac1}), we note that the currents 
become large in the limit $\om_0 \to \pm \om$ if $K<1/2$. Hence the 
perturbative expansion in powers of $U_p$ breaks down when $\om_0$ is close 
to $\pm \om$ \cite{feldman1}. The region of validity of the perturbative 
expansion can be estimated using a RG analysis as discussed below.

% It is interesting to note from Eq. (\ref{idc1}) that the correction to the 
% differential conductance $\Delta G \equiv - \partial I_{bs,dc} /\partial 
% V_{bias} = - q \partial I_{bs,dc} / \partial \om_0$ is positive if $K < 1/2$
% \cite{feldman1}. Namely, the total conductance is larger than $q^2 /(2\pi)$ 
% which is the value for a clean wire with non-interacting electrons.

Eqs. (\ref{idc1}-\ref{iac1}) imply that for the pure pumping case with $\om_0
=0$, $~I_{bs,dc}^{pp} = I_{bs,ac}^{pp} =0$. For a single impurity, therefore,
charge pumping does not occur, whether or not there are interactions between 
the electrons. For the pure bias case with $\om =0$ and $\phi_p =0$, we have 
$~I_{bs,dc}^{pp} + I_{bs,ac}^{pp} \sim U_p^2 \om_0^{2K-1}$. Thus the 
backscattering correction to the conductance given by $- I_{bs,dc} / V_{bias} 
= - q I_{bs,dc} / \om_0$ is proportional to $U_p^2 V_{bias}^{2K-2}$. 

In the presence of both bias and pumping, the correction to the differential 
conductance $\Delta G = - q \partial I_{bs,dc} / \partial \om_0$ grows 
large as $U_p^2 |\om_\pm|^{2K-2}$ for $\om_+$ or $\om_- \to 0$. This is 
consistent with results based on RG calculations \cite{kane,furusaki}. Namely,
the presence of interactions between the electrons effectively makes the 
impurity strength $U_p$ a function of the length scale; this is described by 
the RG equation $dU_p / d \ln L = (1-K) U_p$, to first order in $U_p (L)$.
Hence the value of $U_p (L)$ at a length scale $L$ is related to its value 
$U_p$ defined at a microscopic length scale (say, $\al$) as $U_p (L) = 
(L/\al)^{1-K} U_p$. In our case, the length scale $L$ is set by $v/|\om_+|$ or 
$v/|\om_-|$. The effective impurity strength $U_p (L)$ therefore increases
as $(v/|\om_\pm|)^{1-K} U_p$ for $\om_+$ or $\om_- \to 0$, and the correction
$\Delta G$ grows as $[U_p (L)]^2 \sim U_p^2 |\om_\pm|^{2K-2}$. This divergence
must be cut off when $\Delta G$ becomes of order 1, in units of $q^2/(2\pi)$.
Restoring the appropriate dimensionful quantities, we see that the above RG
analysis and perturbative expansion are valid as long as $U_p /v << (\al 
|\om_\pm| /v)^{1-K}$.

\subsection {Several impurities}

We now consider the case of several impurities located at $x_p$ with the phases
of the oscillating potentials being $\phi_p$. We again define 
$x_{rp}$ and $\phi_{rp}$ as in Eq. (\ref{ir}). The backscattered current can 
be written as $I_{bs}=\sum_p I_{bs}^{pp}+\sum_{p<r} I_{bs}^{pr}$. The dc and 
ac parts of $I_{bs}^{pp}$ are given in the previous subsection. Next, we find
that
\bea I_{bs,dc}^{pr} &=& \frac{qU_pU_r}{4 \pi v^2 \Gamma (2K)}
\left(\frac{\al}{v}\right)^{2K-2} \non \\
&\times& [sgn (\om_+) ~|\om_+|^{2K-1} \cos(2k_F x_{rp}+\phi_{rp}) \non \\
& & + ~sgn (\om_-) ~|\om_-|^{2K-1} \cos(2 k_F x_{rp}-\phi_{rp})], \non \\
& & \label{idc2} \eea
\bea I_{bs,ac}^{pr} &=& \frac{qU_pU_r}{4 \pi v^2 \Gamma (2K) \cos (\pi K)}
\left(\frac{\al}{v} \right)^{2K-2} \cos (2k_F x_{rp}) \non \\
&\times & [sgn (\om_+) ~|\om_+|^{2K-1} \non \\
& & ~\times ~\cos(2\om t + \phi_p + \phi_r + sgn (\om_+) \pi K) \non \\
& & ~+~ sgn(\om_-) ~|\om_-|^{2K-1} \non \\
& & ~\times ~\cos(2 \om t+\phi_p + \phi_r - sgn(\om_-) \pi K)]. 
\non \\
& & \label{iac2} \eea
For the pure pumping case with $\om_0 =0$, we see that $~I_{bs,dc}^{pr} \sim
\om^{2K-1} \sin (2k_F x_{rp}) \sin (\phi_{rp})$, while $I_{bs,ac}^{pr}=0$. 
Eq. (\ref{idc2}) differs from the results given in Ref. \cite{makogon1} due
to the terms involving $2k_F x_{rp}$.

We note that the currents given in Eqs. (\ref{idc1}-\ref{iac1}) and 
(\ref{idc2}-\ref{iac2}) all reverse sign if we change $\om_0 \to - \om_0$
and $x_p \to - x_p$ for all $p$. This is a natural consequence of parity 
reversal, i.e., interchange of left and right.

The dc parts given in Eqs. (\ref{idc1}) and (\ref{idc2}) can be combined to
give a total current $I_{bs,dc} = \sum_p I_{bs,dc}^{pp} + \sum_{p<r} 
I_{bs,dc}^{pr}$,
\bea I_{bs,dc} &=& \frac{q}{8 \pi v^2 \Gamma (2K)} \left(\frac{\al}{v}
\right)^{2K-2} \non \\
&\times& [sgn (\om_+) ~|\om_+|^{2K-1} ~|~\sum_p U_p e^{i(2k_F x_p + \phi_p)}~
|^2 \non \\
&+& sgn(\om_-) ~|\om_-|^{2K-1} ~|~ \sum_p U_p e^{i(2k_F x_p - \phi_p)} ~
|^2 ]. \non \\
& & \label{idctot} \eea
The above expression suggests that the current will be maximized if either 
$2k_F x_p + \phi_p$ or $2k_F x_p - \phi_p$ has the same value for all $p$. 
This means that the potentials in Eq. (\ref{upt}) should be of the form $U_p 
\cos (\om t - 2k_F x_p)$ or $U_p \cos (\om t + 2k_F x_p)$; this describes a 
potential wave traveling to the right or to the left. Such a wave has been 
studied extensively for the case of non-interacting electrons; see Refs. 
\cite{galp,aharony,kash,aizin,flens,maksym,robin,agarwal} and 
\cite{taly1,cunningham,taly2,leek}.

An unusual phenomenon occurs if the interactions are sufficiently
strong, i.e., if $K < 1/2$. If there is no bias, the DC part of the current 
generally goes as $\om^{2K-1}$ which {\it increases} as $\om$ decreases. 
However, it is clear that if $\om$ was exactly zero (time-independent
impurities), then the current would also be zero. These two statements imply 
that the current must be a non-monotonic function of $\om$, and must have at 
least one maximum at some value of $\om$. Finding the location of the maximum 
requires us to go beyond the lowest order perturbative results of this paper.

\begin{figure}[htb]
\begin{center}
\epsfig{figure=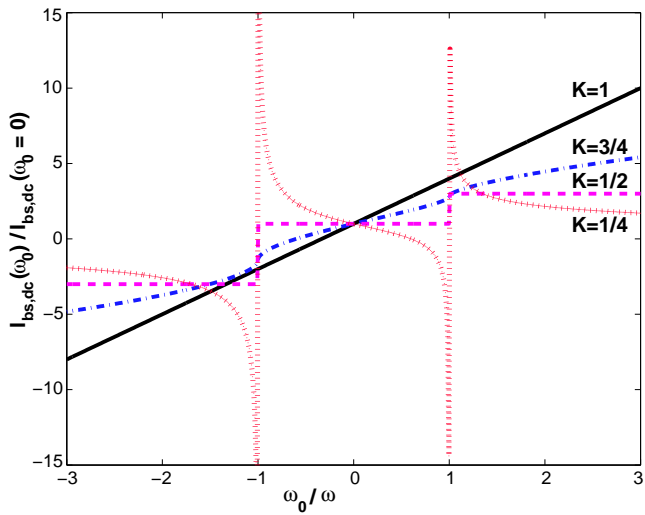,width=8.6cm}
\end{center}
\caption{(Color online) DC part of the backscattered current as a function of
the bias $\om_0$ for several impurities, when $\om_{\pm}$ are small. The red 
(dot), magenta (dash), blue (dash-dot) and black (solid) lines show the 
results for $K=1/4, 1/2, 3/4$ and 1 respectively. We have taken $|\sum_p U_p 
e^{i(2k_F x_p+ \phi_p)}|^2 / |\sum_p U_p e^{i(2k_F x_p- \phi_p)}|^2 = 2:1$.}
\end{figure}

Figure 1 shows the dc part of the backscattered current as a function of the 
applied bias, for a fixed non-zero value of the pumping frequency $\om$,
assuming that $\om$ and $\om_0$ are small. We 
have used the expression in Eq. (\ref{idctot}) to plot the value of $I_{bs,dc}
(\om_0) / I_{bs,dc} (\om_0=0)$ as a function of $\om_0/\om$, for four different
values of the parameter $K=1/4, 1/2, 3/4$ and 1, taking the ratio $|\sum_p U_p
e^{i(2k_F x_p+ \phi_p)}|^2 / |\sum_p U_p e^{i(2k_F x_p- \phi_p)}|^2 = 2:1$ as
an example. For $K=1/4$, the current diverges at $\om_0 = \pm \om$ as 
mentioned above. We also note the linear and piecewise constant dependences 
of the current on $\om_0$ for $K=1$ and 1/2 respectively; this is discussed 
in Subsec. III. E below.

If we relax the assumptions that $\om x_{rp}/v$ and $\om_0 x_{rp}/v$ are 
small, then the exact expressions (up to second order in the impurity 
potentials) for the ac and dc component of the backscattered current are 
given by
\bea I_{bs,dc}^{pr} &=& \frac{qU_pU_r\sqrt{\pi}} {4 \pi v^2 \Gamma(K)} 
\left(\frac{\al}{v}\right)^{2K-2} \left(\frac{2 |x_{rp}|}{v}\right)^{1/2-K} 
\non \\
&\times& [sgn (\om_+) ~|\om_+|^{K-1/2} J_{K-1/2} (|\om_+ x_{rp}|/v) \non \\
& & \times ~\cos (2k_F x_{rp} + \phi_{rp}) \non \\
& & +~ sgn (\om_-) ~|\om_-|^{K-1/2} ~J_{K-1/2} (|\om_- x_{rp}|/v) \non \\
& & \times ~\cos (2 k_F x_{rp}-\phi_{rp})], \non \\ 
& & \label{idc3} \eea

\bea I_{bs,ac}^{pr} &=& \frac{qU_pU_r \sqrt{\pi}}{4 \pi v^2 \Gamma(K) 
\cos(\pi K)} \left(\frac{\al}{v} \right)^{2K-2} \non \\
&\times& \left( \frac{2 |x_{rp}|}{v}\right)^{1/2-K} \cos (2k_F x_{rp}) \non \\
&\times& [ sgn (\om_+) ~|\om_+|^{K-1/2} ~J_{K-1/2} (|\om_+ x_{rp}|/v) \non \\
& & \times ~\cos (2\om t + \phi_p + \phi_r + sgn (\om_+) \pi K) \non \\
& & +~ sgn (\om_-) ~|\om_-|^{K-1/2}~J_{K-1/2}(|\om_- x_{rp}|/v)\non \\
& & \times ~\cos (2 \om t+\phi_p + \phi_r - sgn(\om_-) \pi K) \non \\
& & +~ \{ |\om_+|^{K-1/2}~ J_{1/2-K} (|\om_+ x_{rp}|/v) \non \\
& & ~~~~~-~ |\om_-|^{K-1/2}~ J_{1/2-K} (|\om_- x_{rp}|/v) \} \non \\
& & \times ~\sin(2 \om t + \phi_p +\phi_r) ]. \label{iac3} \eea
The Bessel function $J$ is discussed in the Appendix; using a power
series expansion given there, we can show that Eqs. (\ref{idc3}-\ref{iac3})
reduce to Eqs. (\ref{idc2}-\ref{iac2}) in the limit $\om_\pm x_{rp}/v \to 0$.
We note that the expressions in Eqs. (\ref{idc1}-\ref{iac1}) do not change 
if we relax the assumptions that $\om$ and $\om_0$ are small.

\begin{figure}[htb]
\begin{center}
\epsfig{figure=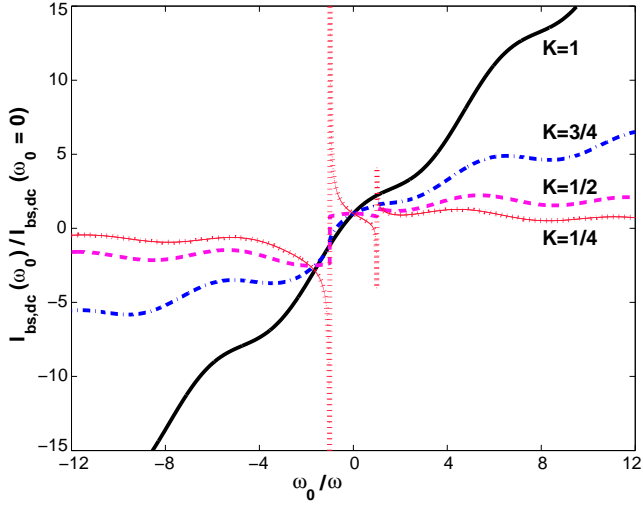,width=8.6cm}
\end{center}
\caption{(Color online) DC part of the backscattered current as a function of
the bias $\om_0$ for two impurities, when $\om_{\pm}$ are not small. The red 
(dot), magenta (dash), blue (dash-dot) and black (solid) lines show the results
for $K=1/4, 1/2, 3/4$ and 1 respectively. We have taken $U_1 = U_2$, $2k_F 
x_{12} = \pi/2$, $\phi_{12} = -\pi/4$, and $\om x_{12}/v = 1$.}
\end{figure}

Figure 2 shows the dc part of the backscattered current as a function of the 
applied bias for the case of two impurities, labeled 1 and 2, taking $U_1 = 
U_2$, $2k_F x_{12} = \pi /2$, $\phi_{12} = - \pi/4$, and $\om x_{12}/v = 1$; 
thus $\om$ and $\om_0$ are not small, in contrast to the case shown in Fig. 1.
We have used the expressions in Eqs. (\ref{idc1}) and (\ref{idc3}) to plot the
value of $I_{bs,dc} (\om_0)/ I_{bs,dc} (\om_0=0)$ as a function of $\om_0/\om$,
for four different values of the parameter $K = 1/4, 1/2, 3/4$ and 1. We see 
some oscillations in Fig. 2 due to the appearance of the Bessel functions in 
Eq. (\ref{idc3}). For $K=1/4$, we again see divergences at $\om_0 = \pm \om$.

\subsection{$K=1$ and 1/2}

We now discuss the special cases $K=1$ and 1/2 where the expressions
for some parts of the currents simplify considerably.

For non-interacting fermions with $K=1$, we find from Eqs. 
(\ref{idc1}-\ref{iac1}) that in the single impurity case, 
\bea I_{bs,dc}^{pp} &=& \frac{qU_p^2}{4 \pi v_F^2} ~\om_0 , \non \\
I_{bs,ac}^{pp} &=& \frac{qU_p^2}{4 \pi v_F^2} ~\om_0 ~\cos (2\om t+2\phi_p). 
\eea
The total current is given by $I = - I_0 + I_{bs,dc}^{pp} + I_{bs,ac}^{pp}$,
\beq I ~=~ \frac{q \om_0}{2\pi} ~\left[ -~ 1 ~+~ \left( \frac{U_p 
\cos(\om t+\phi_p)}{v_F} \right)^2 \right]. \eeq
This is consistent with the fact that the transmission probability across a 
static point-like barrier of height $U$ is $1-(U/v_F)^2$ up to order $U^2$.
For the case of several impurities, we find from Eqs. (\ref{idc2}-\ref{iac2}) 
that
\bea I_{bs,dc}^{pr} &=& \frac{qU_pU_r}{2 \pi v_F^2} ~[~\om_0 ~\cos(2k_F 
x_{rp})~ \cos(\phi_{rp}) , \non \\
& & ~ \quad \quad \quad ~-~ \om ~\sin(2k_F x_{rp}) ~\sin(\phi_{rp}) ~] , \\
I_{bs,ac}^{pr} &=& \frac{qU_pU_r}{2 \pi v_F^2} ~\om_0 ~\cos(2k_F x_{rp})~
\cos(2 \om t+\phi_p+\phi_r) . \non \\
& & \eea 
Note that the dc part of the current is given by a linear combination of 
the pure bias part and the pure pumping part, and it agrees with the
expression given in Eq. (\ref{ir}).

For $K=1/2$, we can obtain the different parts of the currents by taking the
limit $K \to 1/2$ in Eqs. (\ref{idc1}-\ref{iac1}) and (\ref{idc2}-\ref{iac2}).
We find that 
\bea I_{bs,dc}^{pp} &=& \frac{qU_p^2}{8 \pi \al v} ~[sgn (\om_+) ~+~ sgn 
(\om_-)], \non \\
I_{bs,dc}^{pr} &=& \frac{qU_pU_r}{4 \pi \al v} [sgn (\om_+) \cos(2k_F x_{rp}+
\phi_{rp}) \non \\
& & \quad \quad \quad ~+~ sgn(\om_-) \cos(2 k_F x_{rp} - \phi_{rp})], \non \\
I_{bs,ac}^{pp} &=& \frac{qU_p^2}{8 \pi \al v} [(sgn (\om_+) + sgn (\om_-))
\cos(2 \om t + 2 \phi_p ) \non \\
& & \quad \quad \quad ~+~ \frac{2}{\pi} ~\ln | \frac{\om_+}{\om_-} | ~\sin 
(2 \om t + 2 \phi_p )], \non \\
I_{bs,ac}^{pr} &=& \frac{qU_pU_r}{4 \pi \al v} [(sgn (\om_+) 
+ sgn (\om_-)) \cos(2 \om t + \phi_p + \phi_r) \non \\
& & \quad \quad \quad ~+~ \frac{2}{\pi} ~\ln | \frac{\om_+}{\om_-} | ~\sin 
(2 \om t + \phi_p + \phi_r)]. \non \\
& & \eea
Thus the DC parts of the currents do not depend on the precise values of $\om$
and $\om_0$ if they are unequal, and they have a finite discontinuity when 
$\om$ crosses $\pm \om_0$.

To conclude, we see that the dc parts of the currents are linear functions of
$\om_0, \om$ for $K=1$, and are piecewise constant functions of $\om_0, \om$
for $K=1/2$.

\subsection{Extended impurities}

The analysis in Subsec. III. D can be readily generalized to the case where
there is an extended region of oscillating potentials \cite{makogon2}. Let 
us replace the discrete set of potentials given in Eq. (\ref{upt}) by an 
oscillating potential of the following form
\beq U(t) ~=~ \int dx ~U(x) \cos [\om t + \phi (x) ] ~. \eeq
We then see from Eq. (\ref{idctot}) that the dc part of the backscattered
current is given by 
\bea I_{bs,dc} &=& \frac{q}{8 \pi v^2 \Gamma (2K)} \left(\frac{\al}{v}
\right)^{2K-2} \non \\
&\times& [ sgn (\om_+) ~|\om_+|^{2K-1} |\int dx ~U(x) ~e^{i[2k_F x + 
\phi (x)]}|^2 \non \\
&+& sgn (\om_-) ~|\om_-|^{2K-1} |\int dx ~U(x) ~e^{i[2k_F x - \phi 
(x)]}|^2 ]. \non \\
& & \eea
to second order in $U(x)$. For the pure pumping case with $\om_0 =0$, we
find that
\bea I_{bs,dc} &=& - ~\frac{q}{4 \pi v^2 \Gamma (2K)} \left(\frac{\al}{v}
\right)^{2K-2} \om^{2K-1} \non \\
&\times& \int \int dx dx' ~U(x) U(x') ~ \sin [2k_F (x-x')] \non \\
& & \quad \quad ~\times ~\sin [\phi (x) - \phi (x')]. \label{ext} \eea

Eq. (\ref{ext}) implies that the charge pumped per cycle, 
$\Delta Q = (2\pi /\om) I_{bs,dc}$, scales as $\om^{2K-2}$; for $K<1$, this 
grows large in the adiabatic limit $\om \to 0$. In this limit, we saw 
earlier that the effective length-dependent impurity strength diverges 
at small energy scales, which implies that the impurity presents a very large
barrier to the electrons and the transmission coefficient is very small. In
this limit, it has been argued in Refs. \cite{das,sharma} that the pumped 
charge $\Delta Q$ is quantized to be an integer multiple of $q$.

\subsection{Spin-1/2 electrons}

For spin-1/2 electrons in one dimension, the phenomenon of spin-charge 
separation occurs if there are interactions between the electrons. The spin 
and charge degrees of freedom can be separately bosonized 
\cite{gogolin,giamarchi}. The two bosonic theories are characterized by the 
parameters $(K_s , v_s)$ and $(K_c , v_c)$ respectively. For a system with
$SU(2)$ rotational invariance, $K_s = 1$. The ground state
expectation value in Eq. (\ref{expval1}) then takes the form
\bea & & \langle 0| ~\psi_{\si R}^\da (x_p,t') \psi_{\si L} (x_p,t') 
\psi_{\si L}^\da (x_r,t) \psi_{\si R} (x_r,t) ~|0 \rangle \non \\
& & \sim ~\frac{1}{[(x_p-x_r)^2 ~-~ (v_s(t'-t) -i\al)^2]^{1/2}} \non \\
& & \quad \times ~\frac{1}{[(x_p-x_r)^2 ~-~ (v_c(t'-t) -i\al)^2]^{K_c/2}},
\label{expval2} \eea
where $\si = \uparrow , \downarrow$ is the spin label. The appearance of two 
different velocities, $v_s$ and $v_c$, and two different exponents, $1/2$ and 
$K_c /2$, in Eq. (\ref{expval2}) makes the expressions for the backscattered 
current rather complicated. However, we can find the power law of the 
dependence of the currents on the frequencies by a simple scaling argument.
With the approximations made earlier, $\om x_{rp}/v_{s,c}$ and $\om_0 x_{rp}/
v_{s,c} \to 0$, we see that the time dependence has changed from $1/(t' -
t)^{2K}$ in Eq. (\ref{expval1}) to $1/(t' -t)^{K_c+1}$ in Eq. (\ref{expval2}).
The dependences of the backscattered currents on the frequencies therefore 
change from $|\om_0 \pm \om|^{2K-1}$ in the spinless case to $|\om_0 \pm 
\om|^{K_c}$ in the spin-1/2 case. Since $K_c$ is positive in general, the
current no longer diverges as $\om_0 \to \pm \om$.

\section{Discussion}

We have considered the effects of a bias and a number of weak and harmonically
oscillating potentials on charge transport in a Tomonaga-Luttinger liquid. We 
have computed the backscattered current to second order in the amplitudes of 
the potentials. For most of our results, we have assumed the oscillation 
frequency and the bias to be small, but we have relaxed that assumption in 
Eqs. (\ref{idc3}-\ref{iac3}). For our assumption of a Dirac fermion with a 
linear dispersion to be valid for an experimentally realizable system, we 
must of course assume that $\om$ and $\om_0$ are small compared to the band 
width of the electrons.

We find that the
backscattered current is maximized for a traveling potential wave in which the
positions and phases of the oscillating potentials are related in a linear 
way. For spinless electrons, if the interactions are sufficiently repulsive 
with $K < 1/2$, the backscattered current diverges for special values of the 
bias, namely, for $\om_0 \to \pm \om$. For any repulsive interaction, with $K 
< 1$, the correction to the differential conductance diverges for $\om_0 \to 
\pm \om$. Finally, we have pointed out a peculiarity which arises when several
impurities are present and $K<1/2$; namely, the current must in general 
be a non-monotonic function of the pumping frequency when there is no bias.

It would be useful to generalize our results to the case of one or more 
strong impurity potentials, or weak tunnelings between two Tomonaga-Luttinger 
liquids; the technique of bosonization can be used in such situations also.

\section*{Acknowledgments}

A.A. thanks CSIR, India for a Junior Research Fellowship. D.S. thanks Sourin 
Das and Sumathi Rao for stimulating discussions. We thank DST, India for 
financial support under the projects SR/FST/PSI-022/2000 and SP/S2/M-11/2000.

\appendix

\section{Some mathematical formulae}

We need to evaluate integrals of the form
\beq \int_0^\infty d\tau ~\frac{\exp (\pm i \Omega \tau)}{((\tau \pm i \al)^2 
- x^2)^K}, \label{a1} \eeq
where $\Omega = \om_\pm$ and $x=x_{rp}/v$. Eq. (\ref{a1}) can be written as 
the sum of integrals running from $0$ to $x$ and from $x$ to $\infty$. We 
find that there are several integrals from $0$ to $x$ which
cancel each other. One is then left with integrals
running from $x$ to $\infty$ in which one can take the limit $\al \to 0$ 
in the denominator. We then use the following results \cite{gradshteyn}
\bea \int_x^\infty d\tau ~\frac{\sin(\Omega \tau)}{(\tau^2 - x^2)^K}
&=& \frac{\sqrt{\pi}}{2} ~\left(\frac{2x}{\Omega}\right)^{1/2-K} \Gamma (1-K)
\non \\
& & \times ~J_{K-1/2} (\Omega x), \non \\
\int_x^\infty d\tau ~\frac{\cos(\Omega \tau)}{(\tau^2 - x^2)^K}
&=& - ~\frac{\sqrt{\pi}}{2} ~\left(\frac{2x}{\Omega}\right)^{1/2-K} 
\Gamma (1-K) \non \\
& & \times ~Y_{K-1/2} (\Omega x), \eea
where $J$ and $Y$ are Bessel functions of the first and second kind 
respectively. The above equations are valid for $x, \Omega >0$, and 
$0< K < 1$. We then use the power series expansion \cite{abramowitz}
\bea J_\nu (z) &=& \left(\frac{z}{2} \right)^\nu ~\sum_{n=0}^\infty ~(-1)^n
\frac{(z/2)^{2n}}{n! ~\Gamma(n+\nu+1)} , \eea
and the relation
\bea Y_\nu (z) &=& \frac{1}{\sin{(\pi \nu)}}~ [\cos(\pi \nu)~ J_\nu (z) ~-~ 
J_{-\nu} (z) ] \eea
which is valid for non-integer values of $\nu$. 

Finally, the following identities involving Gamma functions are useful
\cite{abramowitz},
\bea \Gamma(1-z) ~\Gamma(z) &=& \frac{\pi}{\sin(\pi z)}, \non \\
\Gamma(z) ~\Gamma(z+1/2) &=& \frac{\sqrt{\pi}}{2^{2z-1}} ~\Gamma(2z). \eea

\end{document}